**Graphene field effect transistors with Niobium contacts and asymmetric transfer characteristics**

Antonio Di Bartolomeo[1], Filippo Giubileo[2], Francesco Romeo[1], Paolo Sabatino[1], Giovanni Carapella[1], Laura Iemmo[1], Thomas Schroeder[3,4], Grzegorz Lupina[3]

[1] Dipartimento di Fisica "E.R. Caianiello" and Centro Interdipartimentale NanoMates, Università degli Studi di Salerno, Fisciano, Salerno, Italy. E-mail: dibant@sa.infn.it

[2] CNR-SPIN, Fisciano, Salerno, Italy

[3] IHP-Microelectronics, Frankfurt (Oder), Germany

[4] BTU Cottbus-Senftenberg, Cottbus, Germany

**Abstract**

We fabricate back-gated field effect transistors using Niobium electrodes on mechanically exfoliated monolayer graphene and perform electrical characterization in the pressure range from atmospheric down to $10^{-4}$ mbar. We study the effect of room temperature vacuum degassing and report asymmetric transfer characteristics with a resistance plateau in the n-branch. We show that weakly chemisorbed Nb acts as p-dopant on graphene and explain the transistor characteristics by Nb/graphene interaction with unpinned Fermi level at the interface.

**1. Introduction**

The superconductor/graphene junction is the ideal platform to study the interaction of Cooper pairs and massless Dirac fermions, a research field that let envisage new fundamental physics as well as innovative device applications. The absence of a bandgap in graphene enables easy formation of ohmic contacts with most superconducting metals. Nonetheless, the band alignment and the vanishing density of states (DOS) of graphene around the Dirac point, as well as defects or chemical residues, may affect current injection and hinder the detection of exotic new phenomena at the superconducting transition. The understanding of the contact formed by a given superconductor with graphene is therefore an important prerequisite to any low temperature investigation.

Heersche et al. [1] used back gated graphene field effect transistors (FETs) with Ti/Al contacts to demonstrate that graphene can support a supercurrent, which is carried either by electrons in the conduction band or by holes in the valence band, and that Josephson effect in graphene is a robust phenomenon. Rickhaus et al. [2] fabricated superconductor-graphene-superconductor (S-G-S) devices based on Niobium (Nb) contacts to study the integer quantum Hall effect and evidenced Andreev processes at the graphene-superconductor interface. Their devices were fabricated with exfoliated graphene on $SiO_2$/p-Si substrate and, when tested as back gated field effect transistors, exhibited asymmetric transfer characteristics with saturation in the p-branch and a field effect mobility around $3000\,cm^2V^{-1}s^{-1}$. Since they observed an exponentially increasing contact resistance for decreasing temperature, to achieve transparent contacts to graphene they used a 4 *nm* Ti layer under Nb. Similarly, Komatsu et al. [3] fabricated S-G-S junctions, with Nb or ReW, to investigate the superconducting proximity effect through graphene. They found that the low transparency of the superconductor/graphene junction is a serious limitation and, only after using an intermediate thin Pd layer (4- to 8-nm thick), they were able to evidence a suppression of the critical current near the graphene charge neutrality point, which was attributed to specular reflection of Andreev pairs at the interface of charge puddles. Mizuno et al. [4] fabricated high-quality suspended monolayer graphene–Niobium nitride (NbN) Josephson junctions and measured a supercurrent at critical temperatures greater than *2 K*. The production of highly transparent graphene–NbN contacts was identified as one of the major experimental challenges. A Ti/Pd intermediate layer was e-beam evaporated on graphene prior to Ar/N$_2$ plasma sputtering of Nb to reduce the damage from energetic ions and improve contact transparency.

To date, Nb is the metal often chosen in the superconductor/graphene investigations for the high critical temperature (9.25 *K*) and for the well-known properties and deposition technology, although it does not seem to establish a good contact with graphene and a thin inter-layer is often added. If and how this extra-layer impacts the physics at the superconducting transition is unclear. In this direction, a deeper understanding of the properties of the Nb/graphene interface and the assessment of its suitability for superconductor/graphene investigations is timely and necessary.

We fabricate graphene field effect transistors that we characterize at room temperature and decreasing pressure with the goal to elucidate specific features of the Nb/graphene contact. We find that gently sputtered Nb forms contacts with specific resistivity (~ $25\,k\Omega\,\mu m$) in the range of that reported for evaporated metals, as Ti or Cr (~1-100 $k\Omega\,\mu m$), and about an order of magnitude higher than the specific contact resistance achieved with strongly chemisorbed metals as Pd or Ni (~0.1-5 $k\Omega\,\mu m$)[5-9]. We distinguish the role of air adsorbates and process residues on the doping of the graphene channel from that of the supporting SiO$_2$ and argue that strain of graphene under the contacts plays an important role in increasing the contact resistance. Furthermore we show that Niobium acts as p-dopant on graphene and that depinning of graphene Fermi level at the contact strongly suppresses the conductance of the transistor in the electron branch. As byproduct, we estimate a lower limit for the workfunction of the Nb film as $\Phi_{Nb} \geq 4.7\,eV$.

## 2. Experimental details

Natural graphite flakes (from NGS Naturgraphit GmbH) were repeatedly cleaved with adhesive tapes and transferred onto SiO$_2$/Si substrates. We used moderately doped p-Si (resistivity 1-10 $\Omega\,cm$) covered by 300 *nm* of thermally grown SiO$_2$ to maximize color contrast for optical identification of few-layer graphene [10]. A short dip (~ 60 *s*) in warm acetone was used to remove glue residuals.

Monolayers and few layers graphene were identified optically and further confirmed by Raman spectroscopy and SEM imaging [11]. Selected graphene monolayers were contacted using electron beam lithography on poly(methyl methacrylate) (PMMA) to define suitable metal patterns, followed by standard lift-off technique. As metal we used 25 *nm* of Nb covered by 75 *nm* of Au deposited in a RF magnetron sputtering system (by MRC Inc.). Au was used to prevent Nb oxidation and to provide enough softness to the pads for the needle contact in a probe station. To effectively remove physisorbed molecules and processing residues as PMMA (see following), the sample was subjected to several hours vacuum degassing at $\sim 3 \times 10^{-7}$ *mbar* before metal deposition.

Respect to electron beam evaporation (EBE), metal sputtering on mono and bilayer graphene often results in significant higher contact resistivity, $\rho_c$ [5][12-13]. The sputtering power is a sensible knob to control $\rho_c$, which can be enhanced by orders of magnitude when the power augments [5-6]. Sputtered atoms can possess large kinetic energy that, when transferred to the graphene layer, can remove carbon atoms and form lattice vacancies. As parts of carbon atoms are milled away, new scattering centers are created and an effective smaller contact area is established, which usually increases $\rho_c$. However, current crowding can make the contact resistivity more a function of graphene width than of area [14] and other factors may contribute the transparency of the interface, so that graphene defectiveness may not necessarily be detrimental. Reduction of the specific contact resistance has been achieved by O$_2$ plasma damaging [15] or by intentional pitting [16] or cutting [17] of graphene in the contact region. Damages may facilitate bonding between carbon atoms and metal, while cuts maximize the contacting of graphene just at its edges, which results in dramatically reduced $\rho_c$ and improved mechanical stability [18].

It is also known that metal grain size and uniformity of the metal film make a difference. Large grains and rough surface go in the direction of reducing the effective contact area and increasing the contact resistance. According to Watanabe et al. [19], the metal microstructure affects the contact resistance to a higher extent than the metal workfunction. A reduced contact resistance has been observed in high vacuum deposition condition, suggesting that also the deposition pressure plays an important role on the final contact resistance [20]. Finally, the sputtering process can lead to tensile strain which can cause non-uniform surface coverage and loss of adhesion, or even severe delamination of the metal from the graphene surface [15]. Strain can affect the lattice constant of graphene as well as the chemical bonding and the charge injection rate between graphene and metal [21-22].

To limit graphene damage during sputtering, we sequentially deposited Nb/Au bilayer at a power density as low as 0.7 $W/cm^{-2}$. The sputtering was made in 99.999% pure Argon at pressure of $4 \times 10^{-3}$ *mbar* with substrate at room temperature. The target purity was 99.98% for Nb and 99.95% for Au. The deposition rates were measured to be 0.3 *nm/s* for Nb and 1.2 *nm/s* for Au. To ensure good purity of sputtered Nb, prior to the deposition, a quite energetic cathode pre-sputtering of 5 *min* at 3.5 $W/cm^{-2}$ was performed. For the used parameters and film thickness, the roughness of the Nb film is estimated to be around 1.5 *nm* [23] and the mean grain size around 20 *nm* [4]. For these characteristics, the Nb film is borderline between the Cr, Fe, Ag films with large grains and large contact resistance and the Co, Ni, Pd films with smaller grain and lower contact resistance reported in Ref. [19].

Furthermore, for the used sputtering conditions, the mean kinetic energy distribution of Nb atoms arriving at the substrate (10 *cm* away from target) is strongly peaked at approximately 1 *eV* [25], that is, at an energy only 3 to 5 times larger than energy typically involved in EBE and significantly lower than the energy of ~7.4 *eV* needed for the formation of carbon vacancies in graphite [26-27]. Therefore limited graphene damage is expected. However, the geometry of the contacts, made of long and thin leads laid on graphene (Figure 1) and the fact that they consist of a bilayer can favor the appearance of tensile stress.

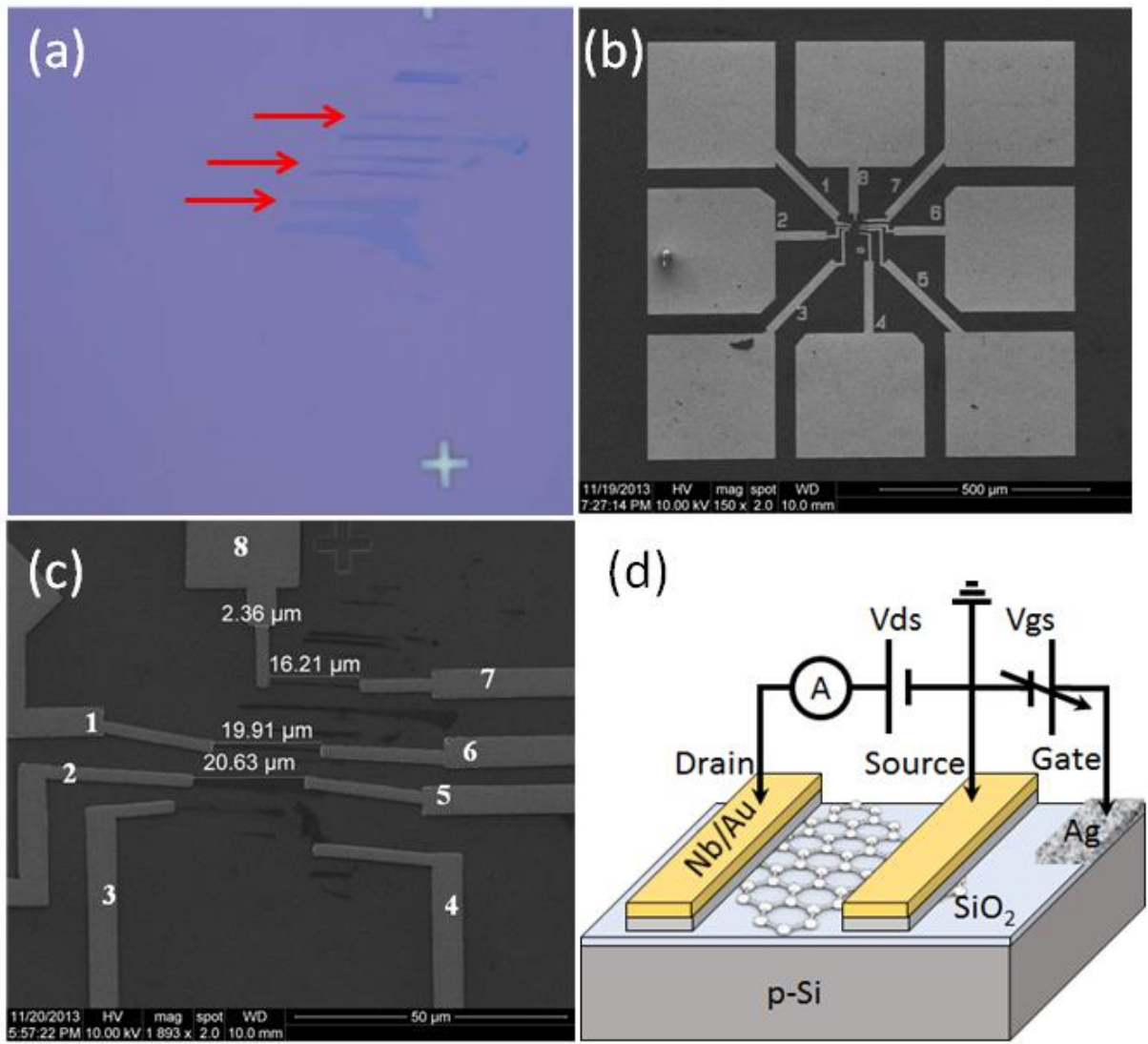

Figure 1 – (a) Optical image of monolayer graphene ribbons under 20× magnification. (b) SEM image of Nb/Au pads and leads used to contact graphene ribbons of (a). (c) Zoom in of the central part of (b) showing contacts on graphene ribbons. The graphene flake between leads 1 and 6 was used in this study. (d) Schematic of the transistor consisting of a layer of graphene used as channel with two Nb leads

functioning as source and drain and the Si substrate acting as back-gate. The 3-terminal measurement consists in monitoring the source-to-drain current $I_{ds}$ under constant bias, $V_{ds}$ = 3 or 5 $mV$, while the gate voltage $V_{gs}$ ranges in the interval (−70 V, 70 V).

Figure 1(a) shows a 20× magnification of few ribbon-shaped graphene monolayers that were contacted in a 2-point configuration. The SEM images of Figure 1(b) and 1(c) show details of the pads and metal leads of the final device. The study presented here is referred to the graphene flake between the leads labelled 1 and 6 in Figure 1(c). The flake is 19.91 $\mu m$ long and 0.79 $\mu m$ large. Similar results were found with the other graphene ribbons.

Electrical measurements were performed at room temperature with the sample under controlled pressure inside a Janis Inc. probe station connected to a Keithley 4200 Semiconductor Parameter Analyzer. The 3-terminals measurement setup is shown in Figure 1(d), which shows also a schematic of the device under study consisting of a layer of graphene used as channel of a FET with two Nb leads functioning as source and drain, kept at constant bias $V_{ds}$ (=3 or 5 $mV$). The Si substrate acts as the transistor back-gate and is swept in the voltage interval (−70 V, 70 V). Higher gate voltages were avoided to prevent oxide damage as stresses at $|V_{gs}| > 80$ V systematically cause either gate leakage or complete oxide breakdown.

The electrical measurements were performed after the sample was kept for given time periods under controlled pressure (down till $2.7 \times 10^{-4}$ mbar). Prior, the sample had been subjected to other measurements which had the effect of stabilizing electric annealing.

## 3. Results and discussion

Figure 2(a) shows the $I_{ds} - V_{gs}$ transfer characteristics of the transistor of Figure 1, at decreasing pressures and at room temperature. These curves are the fixed drain-bias version of the $I_{ds} - V_{ds}$ output characteristics shown in Figure 2(b), whose linear behavior confirms the ohmic nature of the contacts. Figure 2(a) evidences a factor-two gate modulation of the current originating from the vanishing density of states of graphene around the Dirac point ( $D(E) \propto |E|$ ). In air, the device has a clear p-type behavior with a positive Dirac point corresponding to the conductance minimum beyond +50 V. The heavy p-type doping is expected for air exposed graphene and is caused by adsorbed moisture [28] and other chemical residues, such as PMMA not completely removed by acetone during the cleaning process. Both $H_2O$, $O_2$, $NO_2$ molecules [29-31] and PMMA [32-33] are well known p-dopants. Keeping the sample for many hours in vacuum gradually removes physisorbed chemicals and residues [32] [34] and has strong effects on the electrical characteristics of the device. Figure 2 (a) shows that vacuum degassing, even at room temperature, shifts the Dirac point towards negative $V_{gs}$, which corresponds to a gradual transformation of the FET in a device with n-type channel [35]. In our long device, the desorption of acceptor impurities let the $SiO_2$ dielectric to take control of the channel doping, which is transformed in n-type. The n-doping is due to charge transfer from surface states in the $SiO_2$ dielectric to the graphene sheet [36]. After 6 days at low pressure ($\sim 3\times 10^{-4}$ mbar) the device reaches a stable configuration and no appreciable changes are observed over time even with a further lowering of the pressure, indicating that most adsorbates and residues have been removed. Remarkably, simple vacuum degassing at room temperature causes an increase of the mobility (corresponding to steeper V-shaped curves) and a reduction of the minimum conductance. This has been emphasized in Figure 2(c) showing the pressure dependence of the mobility and of the Dirac point $(V_{gs}, I_{ds})$. Long range scattering by charged impurities [37-38], as those related to $O_2$, $H_2O$ or PMMA residues, has a significant impact on carrier motion and on the effective residual carrier density of graphene, $n_0$, which determines the Dirac point conductivity [39]. $n_0$ is due to electron/holes puddles induced by the local potential variations of charged impurities [40] rather than to thermal excitation of carrier above the Fermi level. Therefore, the removal of impurities leads to an increase of the carrier mobility for reduced scattering and to a lowering of $n_0$, i.e. of the minimum conductivity.

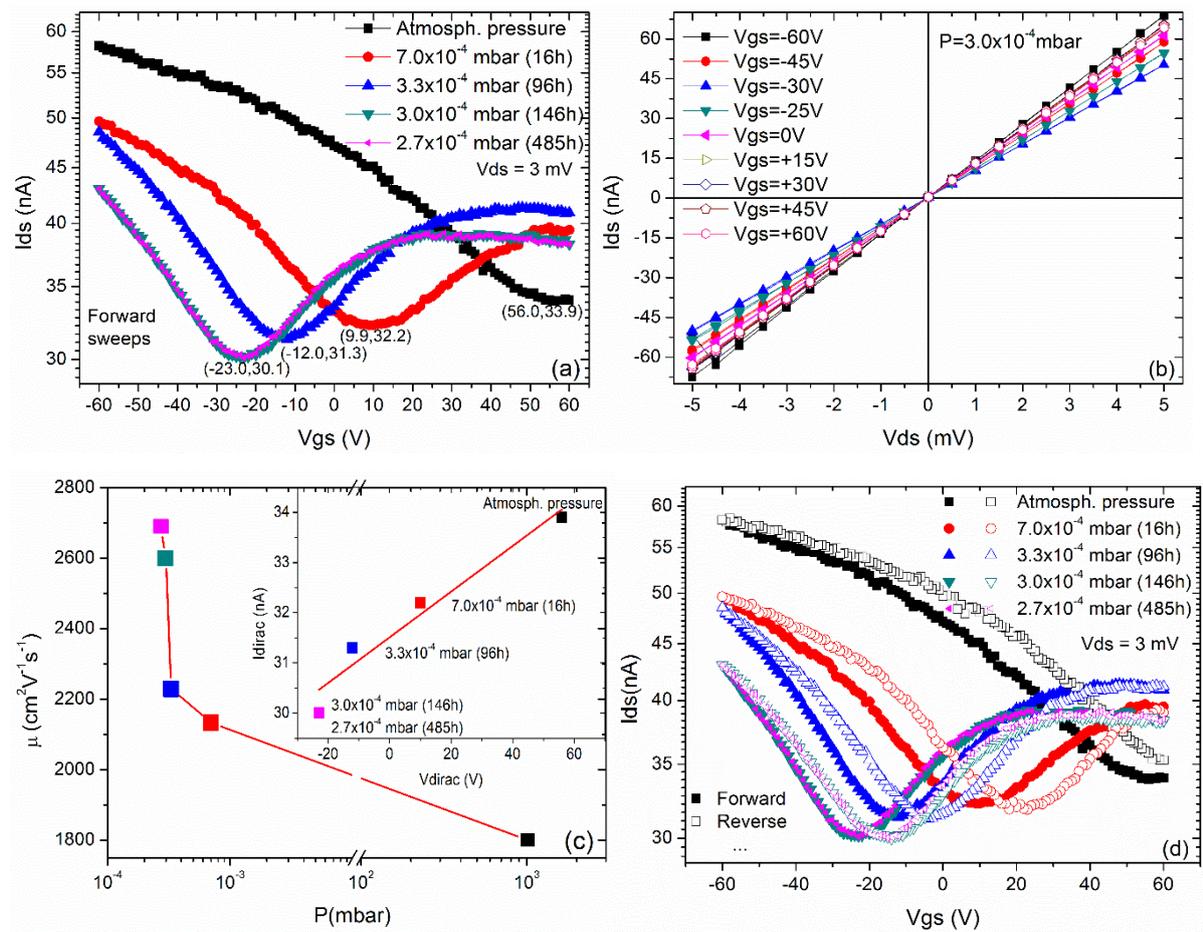

Figure 2 – Electrical characteristics of transistor 1-6 of Figure 1(c). (a) Transfer characteristics $I_{ds} - V_{gs}$ at different pressures and room temperature. (b) Output characteristics $I_{ds} - V_{ds}$ at $3.0\times 10^{-4}$ mbar. (c)

Mobility and $I_{ds}-V_{gs}$ of the Dirac point at different pressures. (d) Full loop $I_{ds}-V_{gs}$ curves at different pressures showing a decreasing hysteresis with higher vacuum.

Another effect of vacuum degassing is the decreased hysteresis of the $I_{ds}-V_{gs}$ curves as can be observed in Figure 2(d) where full loops with forward and reverse $V_{gs}$ sweeps are plotted. Hysteresis is known to be caused by charge trapping at the graphene/SiO$_2$ interface and in the dielectric layer [41]. Removal of impurities, and in particular of H$_2$O which plays a key role in the charge transfer during a $V_{gs}$ sweep [42], obviously results in reduced hysteresis.

With a two-point setup, the total resistance between source and drain of the graphene FET, calculated as $R_{tot}=V_{ds}/I_{ds}$, is the sum of the metal resistance (which is here negligible), the total contact resistance $R_{con}$ and the graphene channel resistance $R_{ch}$ (both $R_{con}$ and $R_{ch}$ can depend on $V_{gs}$ [13]):

$$R_{tot}=R_{con}+R_{ch} \qquad (1)$$

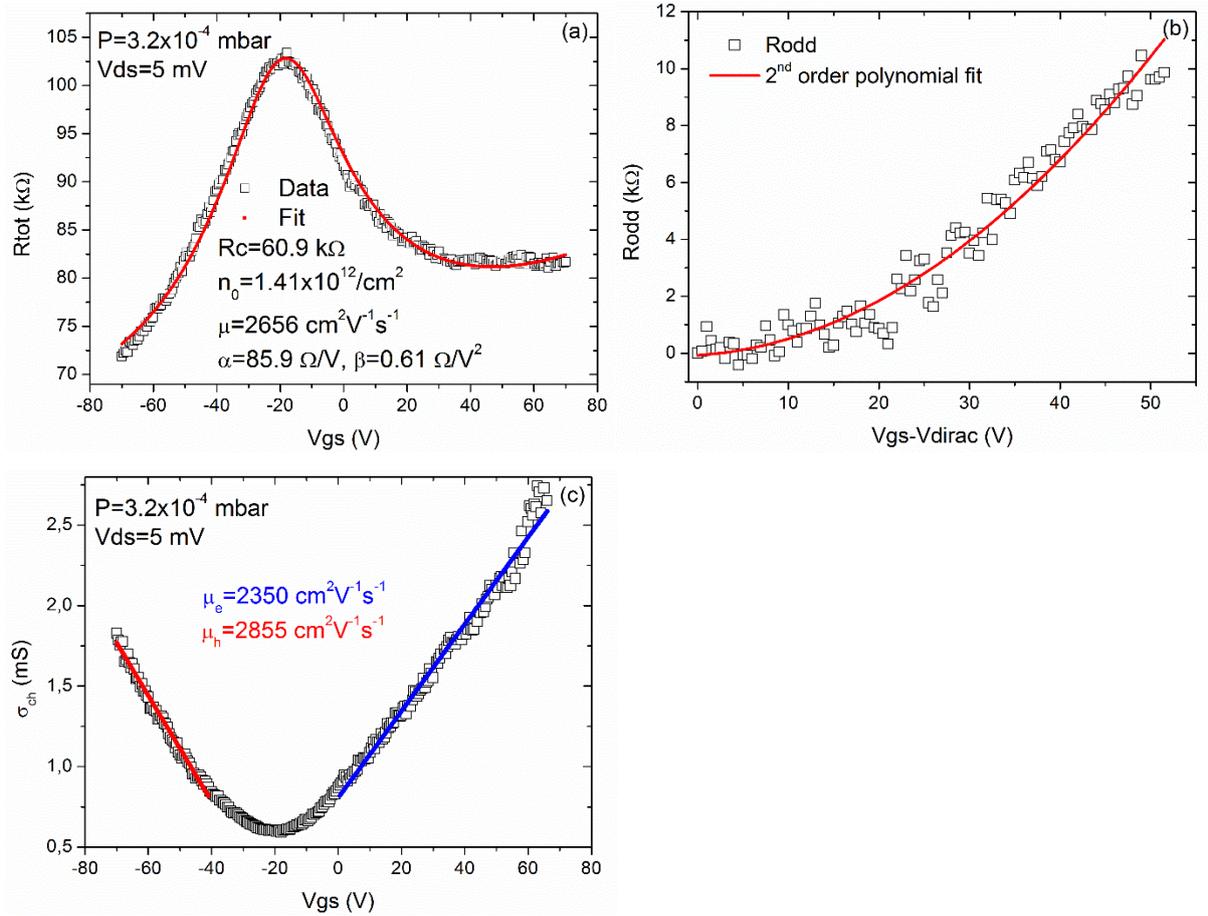

Figure 3 – Electrical characteristics of transistor 1-6 of Figure 1(c). (a) Total resistance $R_{tot}$ as a function of $V_{gs}$ (at $V_{ds}=5$ $mV$ and P=$3.2\times10^{-4}$ $mbar$) and eq. (7) fit (solid line). (b) $R_{odd}$ vs $V_{gs}$ with 2$^{nd}$

order polynomial fit. (c) $\sigma_{ch}$-$V_{gs}$ curve obtained by subtracting the contact resistance contribution and linear fit to estimate electron and hole mobilities.

The measured $R_{tot}$ is shown in Figure 3(a), where two main features can be noted:
1. an asymmetry between the n and the p-branch
2. a resistance plateau in the n-branch, that does not have a counterpart in p-branch

The asymmetry can be characterized by introducing an "odd resistance" $R_{odd}$ defined as

$$R_{odd} = \frac{R(\Delta V_{gs}) - R(-\Delta V_{gs})}{2} \qquad (2)$$

for $\Delta V_{gs} = V_{gs} - V_{dirac} > 0$. $R_{odd}$, shown in Figure 3(b), is positive and has a quadratic dependence on the gate voltage. As we will discuss later, asymmetry is mainly caused by additional pn junctions created by doping at the contacts [43-44]. Then, since $R_{odd}$ is an effect of the contacts and is well fitted by a 2$^{nd}$ order polynomial, we admit a quadratic dependence of $R_{con}$ on $V_{gs}$:

$$R_{con} = R_c + \alpha(V_{gs} - V_{dirac}) + \beta(V_{gs} - V_{dirac})^2 \qquad (3)$$

with $R_c$, $\alpha$ and $\beta$ parameters that we experimentally determine. Such dependence can be easily justified considering the non-linear gate dependence and the spatial inhomogeneity of carrier density in the contact region [45]. We follow the model of Kim et al. [46] to write the conductivity of graphene in the channel as

$$\sigma_{ch} = e n_{tot} \mu \qquad (4)$$

$$n_{tot} = \sqrt{n_0^2 + n^2} \qquad (5)$$

$$n = C_{ox}(V_{gs} - V_{dirac})/e \qquad (6)$$

where $n_{tot}$ is the gate dependent total carrier concentration, $n_0$ is the carrier density at the Dirac point, $n$ is the excess carrier induced by $V_{gs}$, $\mu$ is the mobility (that, in graphene, should be the same for electrons and holes) and $C_{ox} = \varepsilon_{SiO_2}/d = 1.15 \times 10^{-8} \, F/cm^2$ is the capacitance per unit area of the SiO$_2$ layer of thickness $d$.

Using (3)-(6), we express the total resistance as

$$R_{tot} = R_c + \alpha(V_{gs} - V_{Dirac}) + \beta(V_{gs} - V_{dirac})^2 + \frac{L}{W} \frac{1}{e\mu\sqrt{n_0^2 + [(V_{gs} - V_{dirac}) \cdot C_{ox}/e]^2}} \qquad (7)$$

where $L$ and $W$ are the channel length and width, respectively. The fit of (7) to the experimental data, shown in Figure 3(a), yields $R_c \approx 60 \, k\Omega$ (i.e. ~30 $k\Omega$ at each contact, corresponding to a specific contact resistivity $\rho_c = R_c W \approx 24 \, k\Omega \, \mu m$) with less than 10% variation due to the gate dependent terms and a mobility $\mu \approx 2600 \, cm^{-2} V^{-1} s^{-1}$. The fit results quite accurate also in the plateau region at $V_{gs} > 35 \, V$.

Figure 3(c) shows the channel conductivity obtained by eliminating $R_{con}(V_{gs})$, as expressed by (3), from the total measured resistance:

$$\sigma_{ch} = en_{tot}\mu = \frac{1}{R_{tot} - R_{con}}\frac{L}{W} \qquad (8)$$

Following the common practice of using the slope of $\sigma_{ch}$-$V_{gs}$ away from the Dirac point (where $n_0 < n$) to estimate the mobility as

$$\mu = \frac{1}{C_{ox}}\frac{d\sigma_{ch}}{dV_{gs}} \qquad (9)$$

we obtain a hole mobility $\mu_h = 2850\ cm^2V^{-1}s^{-1}$ higher than the electron one $\mu_e = 2350\ cm^2V^{-1}s^{-1}$ in the channel. The average mobility $\mu_{avg} = \frac{\mu_e + \mu_h}{2} = 2602\ cm^2V^{-1}s^{-1}$ is consistent with the value previously estimated. Although the difference $\mu_h - \mu_e$ may be exaggerated by the method which does not take into account the carrier inhomogeneity along the channel (which can be particularly important in the n-branch where a p-n-p structure is formed, as we will see later), this results suggests that there is a mobility contribution to the n and p branch asymmetry. Higher hole mobility has been often measured in graphene transistors [41][47-48]; one plausible explanation is the different scattering cross section for electrons and holes by charged impurities, according to which the massless carriers are scattered more strongly when they are attracted to a charged impurity than when they are repelled from it [49]. In the case under study, after desorption of chemicals and residues, the charged impurities are mainly the positive charges stored in $SiO_2$ dielectric.

The relatively high specific contact resistance as compared to the benchmark of $\rho_c = 100\ \Omega\mu m$ for good contacts can be hardly ascribed to the roughness and grain size of the Nb film, which can be only a minor contributor. Although expected in minimal amount, impurities non-removed by the vacuum degassing and trapped under the metal as well as defects created by the sputtering process are possible additional sources. Nb is an easily-oxidizable metal and could easily react with residuals $O_2$ or $H_2O$ molecules. Indeed we believe that an important source of resistance is the strain induced in graphene by Nb/Au leads. Tensile strain in graphene has been observed to weaken the C-C bond and lower the vibration frequency, thus causing a red shift of the 2D and G bands [50]. Theoretically, it has been pointed out that uniaxial and shear strains may move the Fermi level crossing away from the K points while preserving the cone-like energy dispersion [51]. Scanning tunneling microscopy (STM) studies on graphene have revealed a correlation between local strain and increased tunneling resistance [21]. Uniaxial tensile strain greater than 3% has been proven to cause a dramatic increase of the graphene resistance, an important feature for applications in strain gauge devices [22]. A confirmation of stress in our Nb/Au film, which is transmitted to the graphene underneath, is the observation of metal peeling off which sometimes happened during the fabrication or the measurement process. The strain and the high contact resistance point towards a weak bonding between graphene and Nb, with a likely high

Nb/graphene separation on the atomic scale (> 3 Å). The weak bonding, which in the most severe case can pose practical adhesion problems, favors Fermi level depinning at the contact but preserves the conical electronic structure of graphene.

Another important fact to consider is the Nb-graphene workfunction mismatch. Such a mismatch provokes charge transfer across the interface, forms an interface dipole with an accompanying potential step $\Delta V$ and shifts the graphene Fermi level [9]. The transferred charge, which results in local doping of graphene, is not confined under the contacts but can extend hundreds nanometers in the channel [52]. If the Fermi level is not pinned, the gate voltage is able to further tune the charge density of graphene in the contact region [53-54].

A positive $R_{odd}$ is typically obtained when $\Delta\Phi = \Phi_M - \Phi_{G0} < 0$ where $\Phi_M$ and $\Phi_{G0} = 4.5\,eV$ are the workfunctions of the metal and of the intrinsic graphene, respectively [43]. Metals with higher workfunction than graphene tend to subtract electrons from graphene which becomes locally p-doped. Nb has a workfunction in the range 3.95-4.87 $eV$ [55] and can behave both as acceptor and donor for graphene. According to Leblanc et al. [56] the electron workfunction of pure Nb highly depends on the crystal orientation, with the highest value belonging to the {1,1,0} orientations and the lowest one for the {001} orientations. An appreciable increase of a Nb film workfunction has also been reported for increasing oxygen content [57].

For the measured device the positive $R_{odd}$ and the shape of the transfer characteristic strongly indicates that graphene at the contacts is p-type. A Nb film with lower workfunction and donor behavior can be expected as well and would result in a y-axis specular transfer characteristic with resistance plateau in the p-branch. Actually, as already mentioned, Nb contacted graphene FETs with such characteristic have been reported by Rickhaus et al. [2]. We remark that the sign and amount of doping is a multi-factor effect. Metal-graphene spacing, as affected by strain or imputies and defects, structural modifications, wave function hybridization, etc. may contribute to doping other than workfunction difference. Based on density functional calculations, Giovannetti et al. [9] showed that the formation of the interfacial dipole (experimentally proven by Pi et al. [58]) promotes n-type doping for strongly chemisorbed metals (metal graphene separation ~2Å) with workfunction up to 5.4 $eV$. The same model predicts that p-doping is dominant for most metals when weak chemical interaction, corresponding to high metal-graphene separation (> 3.0-4.0 Å), takes place. The latter is the situation that we have been depicting for the device under study. As a matter of fact, p-doping has been observed even with Ti contacts despite the strong donor character given to it by its low workfunction [14][59].

A qualitative model explaining the whole $I_{ds} - V_{gs}$ behavior of the device is presented in Figure 4. It consider p-doping and Fermi level depinning as expected from the considerations made so far. For $V_{gs} < V_{dirac}$, the band alignment of graphene under the contacts and in the channel is that shown in the inset (1), corresponding to a p-p+-p structure. The p-type conductance of the device is strongly increased by the negative $V_{gs}$ which augments the available DOS both in the channel and under the metal. In the contact region, the metal-graphene workfunction difference and the screening effect of the metal make the gate control of the carrier concentration less effective than in the channel. A minimum of conductance is achieved when, increasing $V_{gs}$, the graphene in the channel reaches the Dirac point, while the contact remains p-type for effect of contact doping. A p-i-p structure (i=intrinsic) is formed as shown in inset (2).

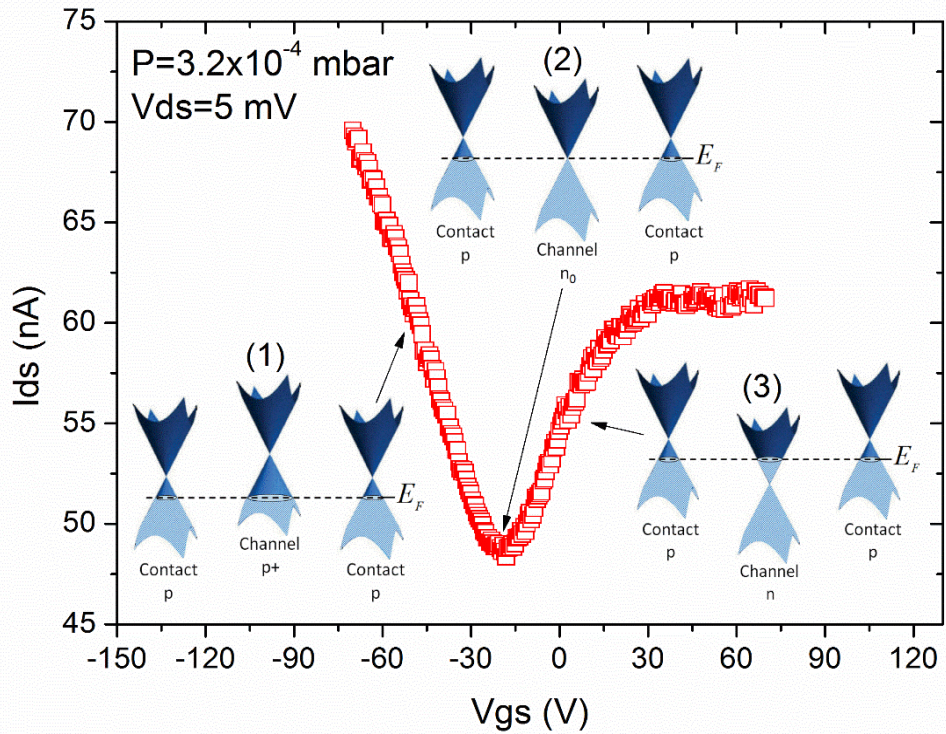

Figure 4 – Fermi level for graphene at the contacts and in the channel accounting for the current behavior as a function of $V_{gs}$. It is assumed that the Fermi level is not pinned at the contacts, where p-doping occurs.

For $V_{gs} > V_{dirac}$, n-doping is induced in the channel and the device becomes a p-n-p structure. The appearance of more resistive p-n junctions [60] in the n-branch as opposed to the p-p+ counterparts in the p-branch is the origin of the observed asymmetry in the V-shaped transfer characteristics. The conductance is initially driven by the DOS of the channel, until the channel doping reaches a level comparable to that of the graphene near the contacts. From this point on, the limitation on the conductance is set by the contacts. The gate voltage tends to shift the Fermi level at the contacts upwards, that is to reduce the hole concentration and increase the contact resistance. The increasing contact resistance counter-balances the decreasing channel resistance and this compensation mechanism results in the observed plateau. In this interpretation, more push of the positive gate voltage would furtherly shift $E_F$ up, till the Dirac point at the contacts, and create a second conductance minimum. A careful look at the curves of Figure 1(a) shows a gradual drop of the current towards high positive $V_{gs}$ which seems to confirm this expectation. Indeed, double conductance minima have been reported in back-gated transistors [53][61].

At the plateau, $R_{tot} \approx 82\,k\Omega$. Assuming that the total resistance is dominated by the contacts, using the measured electron mobility ($\mu_e \approx 2300\,cm^2V^{-1}s^{-1}$), we can estimate a carrier density in graphene of ~ $3.4 \times 10^{12}/cm^2$, which for $n_0 = 1.41 \times 10^{12}/cm^2$ and from eq. (5) corresponds to a gate induced excess electrons of $3.1 \times 10^{12}/cm^2$. Such carrier concentration is obtained when the graphene Fermi level $E_F$ with respect to the conical point is

$$E_F = \hbar v_F \sqrt{\pi n} \approx 0.21\,eV \qquad (10)$$

(here $v_F = 10^6\,m/s$ is the Fermi velocity in graphene). Considering $\Phi_{G0} = 4.5\,eV$ this suggests a Nb workfunction $\Phi_{Nb} \geq 4.7\,eV$. The inequality originates from the fact that, according to Giovannetti et al. model [9], the metal workfunction would be $\Phi_{Nb} = \Phi_{G0} + E_F + e\Delta V$ where the charge dipole voltage $\Delta V > 0$ is a decreasing function of the metal/graphene separation (and should close to zero in the present device).

## 4. Conclusion

We have studied electric properties of graphene FETs with sputtered Nb contacts. We have clarified the role of adsorbates, PMMA residues and underlying SiO$_2$ on the channel doping and distinguished it from the doping at the contacts. We have found that Nb acts as p-dopant but we have clarified that graphene/Nb separation, controlled by stress or other factors, may turn Nb into a donor for graphene. We have shown that the asymmetry observed in the transfer characteristics is naturally explained in terms of doping gradient from contact to channel which gives rise to a p-p$^+$-p structure in the p-branch and to a more resistive p-n-p structure in the n-branch. We have discussed how Fermi level depinning at the contact can limit of electron conductance and create a resistance plateau in the n-branch. We have set a lower limit to the workfunction of Nb as $\Phi_{Nb} \geq 4.7\,eV$.

We have shown that Nb deposited with a low power sputtering forms contacts with graphene of resistivity ($\rho_c \approx 25\,k\Omega \cdot um$) comparable to that achieved with evaporated metals as Ti or Cr and about an order of magnitude higher than that typically achieved with Ni or Pd. We have speculated that a non-negligible contribution to the contact resistance arises from strain. Our finding suggests that further reduction of the Nb/graphene contact resistance is achievable with a careful pre-deposition cleaning, a stress-free design of the metal leads as well as a more gentle deposition, as that of electron beam evaporation. Further improvements, as those needed for the search of new physics or new devices from the superconductor/graphene interface, can be envisaged by using contact resistance reducing techniques, as $O_2$ or ozone treatment or pitting/cutting of graphene, which have been successful in reducing $\rho_c$ of metals as Ni and Pd below the limit of 100 $\Omega \cdot um$.